\documentclass[%
 reprint,
 amsmath,amssymb,
 aps,
]{revtex4-1}

\usepackage{graphicx}
\usepackage{dcolumn}
\usepackage{bm}
\usepackage{amsfonts}
\usepackage{amssymb}

\begin{document}

\title{Testing a Two Field Inflation Beyond the Slow-Roll Approximation}
\author{Kourosh Nozari}
\homepage{knozari@umz.ac.ir}
\author{Kosar
Asadi} \homepage{k.asadi@stu.umz.ac.ir} \affiliation{Department of
Physics, Faculty of Basic Sciences, University of Mazandaran, P.
O. Box 47416-95447, Babolsar, Iran}
\date{\today}

\begin{abstract}
We consider a model of two-field inflation, containing an ordinary scalar
field and a DBI field. We work beyond the slow-roll approximation,
but we assume a separable Hubble parameter. We then derive the
form of potential in this framework and study the spectrum of the
primordial perturbations in details. We also study the amplitude
of the non-Gaussianity of the primordial perturbations both in
equilateral and orthogonal configurations in this setup. We test
the model with recent observational data and find some constraints
on the model parameters. Our study shows that for some ranges of
the DBI parameter, the model is consistent with observation and it
is also possible to have large non-Gaussianity which would be
observable by future improvements in experiments.
\begin{description}
\item[PACS numbers]
98.80.Cq , 98.80.Es
\item[Key Words] Inflation,
Cosmological Perturbations, DBI field, Non-Gaussianity,
Observational Constraints.
\end{description}
\end{abstract}
\maketitle
\section{Introduction}

Inflationary cosmology has became a successful paradigm to
understand the early stage of the universe evolution, with its
advantages of resolving the flatness, horizon and relics problems.
Moreover, during inflation, the vacuum fluctuation of light scalar
fields grow into super-Hubble density perturbations which are
believed to be the origin of the structure formation in the
universe \cite{1,2,3,4,5,6,7,8}. The paradigm of inflation is
essentially related to a quasi-de Sitter universe, a homogeneous
and isotropic universe that expands almost exponentially fast,
with nearly constant event horizon.

Recent observational data have detected a level of scale
dependence in the primordial perturbations \cite{9,10}. Although
there is no direct signal for primordial non-Gaussianity in
observation, however, Planck team has obtained some tight limits
on primordial non-Gaussianity \cite{11}. Some inflationary models
also, predict a level of non-Gaussianity in the primordial
perturbation's mode \cite{12,13,14,15,16,17,18}. In fact, the
primordial non-Gaussianity carries a large amount of information
on the cosmological dynamics deriving the initial inflationary
expansion of the universe. Thus, studying this feature of the
perturbation modes is really an important and interesting issue
and any inflationary model, which can show the non-Gaussianity and
scale dependence of the primordial perturbation is in some sense
more favorable on observational ground.

It is well-known that as a simplest realization, inflation is
derived by a single, slowly-varying scalar field whose potential
energy dominates the universe expansion. However, a single field
inflation model often suffers from fine tuning problems on the
parameters of its potential, such as the mass and the coupling
constant. It has been revealed that when a number of scalar fields
are involved, they can relax many limits on the single scalar
field inflation \cite{19}. Although none of these fields can
result inflation separately, but they are able to work
cooperatively to give an enough long inflationary stage
\cite{20,21,22,23}. Furthermore, there exists good reasons to
believe that inflation might have been driven by more than one
scalar field. First, there are many theories beyond the standard
model of particle physics that involve multiple scalar fields,
such as string theory, grand unified theories, supersymmetry, and
supergravity \cite{24,25,26,27,28,29,30,31,32}. Moreover,
introducing one or more fields may provide attractive features.
For example, hybrid inflation models \cite{33}, which involve two
scalar fields, are able to result sufficient inflationary
expansion and match the observed power spectrum of density
perturbations, while possessing more natural values for their
coupling constants and happening at sub-Planckian field values
\cite{33,34,35}. Furthermore, the single-field case is unusual in
sense since the evolving expectation value of the one field serves
as a clock that determines when the inflation phase ends and the
universe returns to Friedmann-Robertson-Walker expansion
\cite{36}. However, with two or more fields, evolution of one
field can be affected by the fluctuations in the other field(s),
and so, the complex conditions under which inflation ends cannot
be expressed in terms of one degree of freedom (for example some
linear combination of the fields). For these reasons, the issue of
multi-field inflation has became more important recently and many
authors have studied such models \cite{37,38,39,40,41,42,43}.

In this work, we consider an inflation model driven by two scalar
fields, an ordinary scalar field with canonical kinetic term and a
DBI field with non-canonical kinetic term. In fact, one of the
fields describing the inflationary phase of the early universe is
expressed by the radial position of a D3-brane moving in a
"throat" region of a warped compactified space. This proposal is
based on the Dirac-Born-Infeld action \cite{44,45} in which there
is a speed limit upon the motion of the brane, affected by both
its speed and the warp factor of the AdS$_{5}$ throat
\cite{46,47,48}. The effective action in a model with a DBI field
contains a non-standard kinetic term and also a function of the
scalar field besides the potential that is related to the local
geometry of the compact manifold traversed by the D3-brane
\cite{46}. Furthermore there is an interesting phenomenological
feature in the DBI inflation, that it results non-Gaussian
signatures in the Cosmic Microwave Background \cite{49,50}.
In Ref. \cite{Lang08} the authors have studied a multi-field DBI
inflation. They have shown that adiabatic and entropy modes
in this setup propagate with the same effective sound speed and so
get amplified at the sound horizon crossing. They have also found that for
small sound speed, the amplitude of the entropy modes is
much larger than the amplitude of the adiabatic modes.
This feature can strongly affect both the observable curvature power
spectrum and the amplitude of non-Gaussianities without
changing the shape relative to the single field DBI case.
The authors of Ref. \cite{Lang09} have studied a multi-field DBI
inflation by considering some bulk fields present in generic flux
compactification. They have investigated also the consequences of the bulk
form fields on scalar cosmological perturbations, both at linear
and non-linear levels. As an important result, they have shown that
the terms due to the fluctuations of the U(1) gauge field confined
on the brane can be compensated exactly by the terms arising from
the coupling between the bulk forms and the brane position scalar
fields in the second and third order actions. Vector-type perturbations
associated with the U(1) gauge field confined on the D3-brane are studied
in this framework too, in order to see possible amplification of their
quantum fluctuations. The gravitational wave constraints on DBI
inflation has been studied in this setup when there is a transfer from
entropy into adiabatic perturbations. As a result, an ultra-violet DBI
multi-field scenario is compatible with data in contrast with the single
field case which is in tension with data. In Ref. \cite{Miz09} the leading
order connected four-point function and the full quantum
trispectrum of the primordial curvature perturbation are
computed in multi-field DBI inflation models. They have shown that in
the squeezed and counter-collinear limits the consistency relations
hold as in single field models. They have shown also that adiabatic,
mixed and purely entropic contributions have different momentum
dependence in this setup. So the trispectrum has the potential
to distinguish between the multi-field and single field DBI inflation
models if the amount of the transfer from the entropy perturbations to
the curvature perturbation is significantly large.

In Ref. \cite{51} the authors have considered an inflationary model
driven by an ordinary scalar field and a DBI field. They have
studied the evolution of the non-adiabatic pressure perturbation
during inflation phase in this setup. Their analysis is based on
the double quadratic potential \cite{52} in the form
$V(\phi,\chi)=\frac{1}{2}m_{\phi}^{2}(\phi^{2}+\Gamma \chi^{2})$.
They have shown also that the evolution of the non-adiabatic
pressure perturbation and also its final amplitude depend strongly
on the kinetic terms. Here we consider neither slow roll nor a
separable potential; instead, to compare our inflationary model
with observation, we suppose just a separable Hubble parameter. We
study the spectrum of the primordial modes of perturbations in
details. Non-Gaussian features of perturbations distribution
parameterized by the quantity $f_{NL}$ characterizing the
bispectrum and generated by the evolution of scalar perturbations
on super-Hubble scales are also treated carefully. We emphasize
that our analysis is done beyond the slow-roll approximation, but
we adopt a separable Hubble parameter. In other words, since
slow-roll condition can be temporarily violated during inflation
(for example if fields start to decay during inflation as
in staggered/cascade inflation \cite{53,54,55,56,57}, if a bump in
the potential is encountered, and it is necessarily violated at
the end of inflation and during reheating), we go beyond this
approximation and our strategy is based on the first order
Hamilton-Jacobi formalism developed by Salopek and Bond in Ref.
\cite{58}, which allows us to express inflationary parameters in
the model, without having to focus on a slow-roll regime (one can
see \cite{59} for application of this formalism to the
single-field case).

There are important parameters in an inflationary model such as
the tensor-to-scalar ratio and the scalar spectral index which
express the main properties of the cosmological perturbations.
Therefore, confrontation of the inflation model with observation
and constraining the model's parameters is an important task
toward realization of more natural models. The constraints $r <
0.13$ and $n_s = 0.9636 \pm 0.0084$ is obtained from the combined
WMAP9+eCMB+BAO+H$_{0}$ data \cite{60}. The conditions expressed by
the joint Planck2013+WMAP9+BAO data are as $r < 0.12$ and $n_s =
0.9643 \pm 0.0059$ \cite{61}. Recently, the Planck collaboration
released the constraints $r < 0.099$ and $n_s=0.9652 \pm 0.0047$
from Planck TT, TE, EE+low P+WP data \cite{9,10,11}. Thus, in
order to compare our model with observational data, we study the
behavior of the tensor-to-scalar ratio versus the scalar spectral
index in the background of the Planck TT, TE, EE+low P dataset and
obtain some constraints on the parameters space of the model.
Furthermore, we study numerically the non-Gaussianity feature of
the model by studying the behavior of the orthogonal configuration
versus the equilateral configuration in the background of the
observational data. We show that for some ranges of the DBI
parameter, our model is consistent with observation and it is also
possible to have large non-Gaussianity. We note that large
non-Gaussianities would be observable by future improvements in
experiments and in this respect this would be an important result
in our study.

The paper is organized as follows: after introducing the setup in
Section II, we investigate the linear perturbation of the model in
section III. By expanding the action up to the second order in
perturbation, we obtain the two-point correlation functions which
results in the amplitude of the scalar perturbation and its
spectral index. Also, by studying the tensor part of the perturbed
metric, we obtain the tensor perturbation and its spectral index
as well. In order to investigate non-linear perturbation in the
model, in section IV, the action is expanded up to the cubic order
in perturbation. To study the non-Gaussian modes of the primordial
perturbations, we consider the three-point correlation functions.
In this section, the amplitude of the non-Gaussianity is obtained
in the equilateral and orthogonal configurations and in $k_1 = k_2
= k_3$ limit. In other words, we focus on the possibility to
obtain a large level of non-Gaussianity beyond the slow-roll
inflation and derive some conditions to have large
non-Guassianity. In section V, we test our two field inflationary
model in confrontation with the recently released observational
data. Finally, we conclude in section VI.

\section{The Model}
We consider a model of inflation driven by two minimally coupled
scalar fields, a scalar field with a canonical kinetic term and a
DBI field, described by the action
\begin{eqnarray}\label{1}
S=\int\sqrt{-g}\Bigg[\frac{M_{pl}^{2}}{2}{\cal{R}}-\frac{1}{2}\partial_{\mu}{\phi}\partial^{\mu}{\phi}\hspace{2cm}\nonumber\\
-f^{-1}(\chi)(1-\gamma^{-1})-V(\phi,\chi)\Bigg]d^{4}x
\,,\hspace{0.5cm}
\end{eqnarray}
where, ${\cal{R}}$ is the Ricci scalar and $M_{pl} = (8\pi
G)^{-1/2}$ is the reduced Planck mass. $\phi$ is the ordinary
scalar field and $\chi$ is the DBI field whereas, $V$ is the
potential of the model which is a function of both fields.
$\gamma=\frac{1}{\sqrt{1-f(\chi)\partial_{\alpha}{\chi}\partial^{\alpha}{\chi}}}$
is the warp factor describing the shape of the extra dimensions
and $f^{-1}(\chi)$, which is the inverse brane tension, is related
to the geometry of the throat in the original DBI framework
\cite{46,47}.

We consider a spatially flat Friedmann-Robertson-Walker spacetime
\begin{eqnarray}\label{2}
ds^{2} = -dt^{2} + a^{2}(t)\delta_{ij}dx^{i}dx^{j},
\end{eqnarray}
where $a(t)$ is the scale factor. The equations of motion for both
fields are given by
\begin{eqnarray}\label{3}
\ddot{\phi}+3H\dot{\phi}+V_{,\phi}=0\,,
\end{eqnarray}
\begin{eqnarray}\label{4}
\ddot{\chi}+3H\gamma^{-2}\dot{\chi}
+\frac{1}{2}f_{,\chi}f^{-2}\bigg[1-3\gamma^{-2}+2\gamma^{-3}\bigg]\nonumber\\
+\gamma^{-3}V_{,\chi}=0\,\,,
\end{eqnarray}
and Einstein's equations give
\begin{eqnarray}\label{5}
H^2=\frac{1}{3M_{pl}^2}\Bigg(\frac{1}{2}\dot{\phi}^2+\frac{1}{f(\chi)}(\gamma-1)+V\Bigg)\,,
\end{eqnarray}
\begin{eqnarray}\label{6}
-2\dot{H}=\dot{\phi}^2+\gamma\dot{\chi}^2\,\,.
\end{eqnarray}
where a dot denotes the derivative with respect to cosmic time
$t$, while `` , "  represents the derivative with respect to the
scalar field. $H=\frac{\dot{a}}{a}$ is the Hubble parameter.

Following the notation of slow-roll parameters defined by
$\epsilon\equiv-\frac{\dot{H}}{H^{2}}$ and
$\eta\equiv-\frac{1}{H}\frac{\ddot{H}}{\dot{H}}$, we find these
parameters in our setup as follows
\begin{eqnarray}\label{7}
\epsilon=\frac{M_{pl}^2 (3\dot{\phi}^2+3\gamma
\dot{\chi}^2)}{\dot{\phi}^2+2f^{-1}(\gamma-1)+2V}\,,
\end{eqnarray}
\begin{eqnarray}\label{8}
\eta=\frac{-2\Big(\dot{\phi}\ddot{\phi}+\gamma\dot{\chi}\ddot{\chi}+\frac{1}{2}\gamma^3\Big(f\ddot{\chi}+
\frac{f_{,\chi}}{2}\dot{\chi}^2\Big)\dot{\chi}^3\Big)}{H\Big(\dot{\phi}^2+\gamma\dot{\chi}^2\Big)}\,\,.
\end{eqnarray}
It is important to note that we have not used the slow-roll
conditions in the calculation of $\epsilon$ and $\eta$. Up to now,
we obtained main equations of this inflationary setup. In the next
section, in order to test this inflationary model, we study the
linear perturbation of the primordial fluctuations. To this end,
we calculate the spectrum of perturbations produced due to quantum
fluctuations of the fields about their homogeneous background
values.

\section{Linear Perturbations}

Now, we study linear perturbations of the two field model
introduced in the previous section. To this end, we expand the
action up to the second order of fluctuations within the ADM
formalism in which we can eliminate one extra degree of freedom of
perturbations at the beginning of the calculation by choosing a
suitable gauge.

The space-time metric in the ADM formalism is
\begin{equation}\label{9}
ds^{2}=-N^{2}dt^{2}+h_{ij}\big(dx^{i}+N^{i}dt\big)\big(dx^{j}+N^{j}dt\big)\,,
\end{equation}
with $N$ being the lapse function and $N_i$ the shift vector. By
expanding the lapse function $N$ and the shift vector $N_{i}$, as
$N = 1+2\Phi$ and $N^i = \delta^{ij}\partial_{j}B$, the general
perturbed form of the metric will be obtained. There is no need to
compute $N$ or $N_i$ up to the second order, since the second
order perturbation is multiplied by a factor which is vanishing
using the first order solution. We also note that, the
contribution of the third order term vanishes. This is because it
is multiplied by a constraint equation at the zeroth order obeying
the equations of motion. $h_{ij}$ is written as
$h_{ij}=a^2[(1-2\Psi)\delta_{ij}+2{\cal{T}}_{ij}]$, where $\Psi$
is the spatial curvature perturbation and ${\cal{T}}_{ij}$ is a
spatial shear 3-tensor which is symmetric and also traceless. So,
the above perturbed metric (\ref{9}) becomes
\begin{eqnarray}\label{10}
ds^2=-(1+2\Phi)dt^2+2a(t)B_{,i}dtdx^i \hspace{2cm}\nonumber\\
+a^2(t)[(1-2\Psi)\delta_{ij}+2{\cal{T}}_{ij}]dx^{i}dx^{j}\,.\hspace{0.5cm}
\end{eqnarray}

In what follows, to study the scalar perturbations, we choose the
uniform-field gauge, $\delta\phi=0$ (which fixes the
time-component of a gauge-transformation vector $\xi^{\mu}$), and
the gauge ${\cal{T}}_{ij}=0$. Finally, by considering the scalar
part of the perturbations at the linear level, the perturbed
metric can be rewritten as
\begin{eqnarray}\label{11}
ds^{2}=-\big(1+2\Phi\big)dt^{2}+2a(t)B_{,i}dx^{i}dt\hspace{2cm} \nonumber\\
+a^{2}(t)\big(1-2\Psi\big)\delta_{ij}dx^{i}dx^{j}\,.\hspace{1cm}
\end{eqnarray}
By expanding the action (\ref{1}) up to the second order in the
perturbations, we obtain
\begin{eqnarray}\label{12}
\emph{S}_2=\int dt d^3x a^3
\bigg[-3M_{pl}^2\dot{\Psi}^2+\frac{M_{pl}^2}{a^2}(2\dot{\Psi}-2H\Phi)\partial^2B \nonumber\\
-2\frac{M_{pl}^2}{a^2}\Phi\partial^2\Psi+6M_{pl}^2H\Phi\dot{\Psi}+
\Big(\frac{1}{2}\dot{\phi^2}+\frac{1}{2}\gamma\dot{\chi}^2\hspace{0.75cm}\nonumber\\
+\frac{1}{2}f\gamma^3\dot{\chi}^4-3M_{pl}^2H^2\Big)\Phi^2+\frac{M_{pl}^2}{a^2}(\partial{\Psi})^2\bigg]\,.\hspace{1cm}
\end{eqnarray}
Variation of the action with respect to $N$ and $N_i$ yields the
following constraints,
\begin{eqnarray}\label{13}
\frac{1}{a^2}\partial^2B=3\dot{\Psi}-\frac{1}{a^2H}\partial^2{\Psi}\hspace{4cm}\nonumber\\
+\frac{1}{M_{pl}^2H}\bigg(\frac{1}{2}\dot{\phi}^2+\frac{1}{2}\gamma\dot{\chi}^2
+\frac{1}{2}f\gamma^3\dot{\chi}^4-3M_{pl}^2H^2\bigg)\Phi\,,\hspace{0.5cm}
\end{eqnarray}
\begin{eqnarray}\label{14}
\Phi=\frac{1}{H}\dot{\Psi}\,.
\end{eqnarray}
Making use of the above results and doing some integrations by
parts, the second order action takes the form
\begin{eqnarray}\label{15}
\emph{S}_2=\int dt d^3x a^3{\cal{W}}
 \bigg[\dot{\Psi}-\frac{c_s^2}{a^2}({\partial{\Psi}})^2\bigg]\,,
\end{eqnarray}
where
\begin{eqnarray}\label{16}
{\cal{W}}=\frac{{\dot{\phi}+\gamma\dot{\chi}^2+f\gamma^3\dot{\chi}^4}}{2H^2}\,,
\end{eqnarray}
and the sound speed, $c_s^2$, is defined as
\begin{eqnarray}\label{17}
c_s^2=\frac{M_{pl}^2(\dot{\phi}^2+\gamma\dot{\chi}^2)}{\dot{\phi}+\gamma\dot{\chi}^2+f\gamma^3\dot{\chi}^4}\,.
\end{eqnarray}
Now, in order to obtain the quantum perturbations of $\Psi$, one
can vary the action (\ref{15}) and find the equation of motion of
the curvature perturbation, $\Psi$, as follows
\begin{eqnarray}\label{18}
\ddot{\Psi}+\Bigg(3H+\frac{\dot{\cal{W}}}{{\cal{W}}}\Bigg)\dot{\Psi}+c_s^2\frac{k^2}{a^2}\Psi=0\,.
\end{eqnarray}
The solution of this equation, up to the lowest order in the
slow-roll variables gives
\begin{eqnarray}\label{19}
\Psi=\frac{iHe^{-ic_s^2k\tau}}{2(c_sk)^{3/2}\sqrt{\cal{W}}}(1+ic_sk\tau)\,.
\end{eqnarray}

By computing the two point correlation function in our setup, we
are able to study the power spectrum of the curvature
perturbation. The two-point correlation function of curvature
perturbations can be derived by obtaining the vacuum expectation
value at the end of inflation
\begin{equation}\label{20}
\langle 0|\Psi(0,\textbf{k}_{1})\, \Psi
(0,\textbf{k}_{2})|0\rangle=\frac{2\pi^{2}}{k^3}{\cal{A}}_{s}\big(2\pi\big)^{3}\delta^{3}\big(\textbf{k}_{1}+\textbf{k}_{2}\big)\,,
\end{equation}
where ${\cal{A}}_{s}=\frac{H^2}{8\pi^2{\cal{W}}c_s^3} $ is the
power spectrum of the scalar perturbations and is evaluated at
$c_{s}k=aH$ ($k$ is the comoving wave number). Its spectral index
can be derived as follows
\begin{eqnarray}\label{21}
n_s-1=-2\epsilon-\frac{1}{H}\frac{d}{dt}\ln{c_s}-\frac{1}{H}\frac{d}{dt}\ln{\epsilon}\,.
\end{eqnarray}
Let us now proceed further to obtain power spectrum of the
gravitational waves in this model. We study the tensor
perturbations of the form
\begin{eqnarray}\label{22}
ds^2=-dt^2+a^2(t)(\delta_{ij}+h^{TT}_{ij})dx^{i}dx^{j}
\end{eqnarray}
where $h^{TT}_{ij}$ is transverse and traceless. It is known that
the $h^{TT}_{ij}$ can be written in terms of the two polarization
modes, as $h^{TT}_{ij}=h_{+}e^{+}_{ij}+h_{\times}e^{\times}_{ij}$.
We choose the normalization for the two matrices such that, in
Fourier space,
\begin{eqnarray}\label{23}
e^{(+)}_{ij}(k)e^{(+)}_{ij}(-k)^*=2\,,
\end{eqnarray}
\begin{eqnarray}\label{24}
e^{(\times)}_{ij}(k)e^{(\times)}_{ij}(-k)^*=2\,,
\end{eqnarray}
and
\begin{eqnarray}\label{25}
e^{(+)}_{ij}(k)e^{(\times)}_{ij}(-k)^*=0\,.
\end{eqnarray}
In this case the second-order action for the gravitational waves
can be expressed as
\begin{eqnarray}\label{26}
\emph{S}_T=\int
dtd^{3}xa^3{\cal{W}}_{T}\Bigg[\dot{h}^2_{(+)}-\frac{c_T^2}{a^2}(\partial{h_{(+)}})\hspace{1cm}\nonumber\\
+\dot{h}^2_{(\times)}-\frac{c_T^2}{a^2}(\partial{h_{(\times)}})\Bigg]\,,
\end{eqnarray}
where ${\cal{W}}_{T}=\frac{M_{pl}^2}{4}$ and $c^2_{T}=1$.

The power spectrum of tensor perturbations, which is obtained by
strategy as performed for the scalar perturbations, is as follows
\begin{eqnarray}\label{27}
{\cal{A}}_T=\frac{H^2}{2\pi^2{\cal{W}}_T}\,,
\end{eqnarray}
which results the following spectral index of gravitational waves
\begin{eqnarray}\label{28}
n_T=\frac{d\ln{\cal{A}}_T}{dN}=-2\epsilon\,.
\end{eqnarray}
Another important inflationary parameter is the tensor-to-scalar
ratio, which in this model takes the following form
\begin{eqnarray}\label{29}
r=\frac{{\cal{A}}_T}{{\cal{A}}_s}=16c_s\epsilon\,.
\end{eqnarray}
This is the consistency relation in this model. Up to this point,
we have calculated the primordial fluctuations in linear order. In
what follows, we explore the non-Gaussianity of the density
perturbations by studying the nonlinear perturbations.

\section{Nonlinear Perturbations and Non-Gaussianity}

Now, we study the non-Gaussianity of the primordial density
perturbation which is another important aspect of an inflationary
model. It follows that to compute the amount of non-Gaussianity in
specific inflation models we need to go beyond the linear-order
perturbation theory. Since the two-point correlation function of
the scalar perturbations gives no information about the
non-Gaussianity of perturbations distribution, one has to study
higher order correlation functions. The most appropriate
correlation function to study the non-Gaussian feature of the
primordial perturbations is the three-point correlation function.
In order to calculate the three-point correlation function, the
action (\ref{1}) should be expanded up to the cubic order in the
small fluctuations around the homogeneous background solution.
Note that cubic terms obtained in this manner result in a change
both in the ground state of the quantum field and also
non-linearities in the evolution. After expanding the action
(\ref{1}) up to the third order in perturbation, the next step is
to eliminate the perturbation parameter $\Phi$ in the expanded
action. By introducing an auxiliary field ${\cal{Q}}$ satisfying
the following relation
\begin{eqnarray}\label{30}
B=-\frac{1}{H}\Psi+\frac{a^2}{M_{pl}^2}{\cal{Q}}\,,
\end{eqnarray}
and
\begin{eqnarray}\label{31}
\partial^2{\cal{Q}}={\cal{W}}\dot{\Psi}\,,
\end{eqnarray}
the third order action up to the leading order can be written as
\begin{eqnarray}\label{32}
S_3=\int dt d^3x
\bigg[-\frac{3M_{pl}^2a^3}{c_s^2}\epsilon\Big(\frac{1}{c_s^2}-1\Big)\Psi\dot{\Psi}^2 \hspace{2cm}\nonumber\\
+aM_{pl}^2\epsilon\Big(\frac{1}{c_s^2}-1\Big)\Psi(\partial\Psi)^2+
\frac{a^3M_{pl}\epsilon}{Hc_s^2}\Big(\frac{1}{c_s^2}-1-2\frac{\lambda}{\Sigma}\Big)\dot{\Psi}^3\nonumber\\
-2\frac{a^3\epsilon}{c_s^2}\dot{\Psi}(\partial_i\Psi)(\partial_i{\cal{Q}})\bigg]\,,\hspace{1cm.}
\end{eqnarray}
where the parameters $\lambda$ and $\Sigma$ are defined as follows
\begin{eqnarray}\label{33}
\lambda=\frac{f\dot{\chi}^4}{4(1-f\dot{\chi}^2)^\frac{3}{2}}+\frac{f\dot{\chi}^6}{3(1-f\dot{\chi}^2)^\frac{5}{2}}\,,
\end{eqnarray}
\begin{eqnarray}\label{34}
\Sigma=\frac{1}{2}\big(\dot{\phi}^2+\gamma\dot{\chi}^2+f\gamma^3\dot{\chi}^4)\,.
\end{eqnarray}

Now, having obtained the third order action, we can proceed to
study the non-Gaussianity of the primordial perturbations by
evaluating the three point correlation functions. In order to
calculate the three point correlation function, we use the
interaction picture where $H_{int}$, the interacting Hamiltonian,
is equal to the lagrangian of the cubic action. The vacuum
expectation value of the curvature perturbation for the
three-point operator in the conformal time interval between the
beginning of the inflation, $\tau_{i}$, and the end of the
inflation, $\tau_{f}$, is given by the following expression
\cite{62,63,64}
\begin{eqnarray}\label{35}
\langle \Psi(\textbf{k}_{1})\, \Psi (\textbf{k}_{2})\, \Psi
(\textbf{k}_{3})\rangle=\hspace{4.5cm}\nonumber\\
-i\int_{\tau_{i}}^{\tau_{f}}d\tau\,a\,\langle
0|[\Psi(\textbf{k}_{1})\,\Psi (\textbf{k}_{2})\,
\Psi(\textbf{k}_{3})\,,\, H_{int}]|0\rangle\,.\hspace{0.75cm}
\end{eqnarray}
Solving the integral in the above equation results the following
three-point correlation function of the curvature perturbation in
the Fourier space
\begin{eqnarray}\label{36}
\langle \Psi(\textbf{k}_{1})\, \Psi (\textbf{k}_{2})\, \Psi
(\textbf{k}_{3})\rangle=\hspace{4.5cm}\nonumber\\
\big(2\pi\big)^{3}\delta^{3}\big(\textbf{k}_{1}+\textbf{k}_{2}+
\textbf{k}_{3}\big){\cal{A}}_s^2\,\,{\cal{F}}_{\Psi}({k}_{1},{k}_{2},{k}_{3})\,,\hspace{0.75cm}
\end{eqnarray}
where ${\cal{A}}_s$ is the power spectrum of perturbation some
time after the Hubble radius crossing, and
\begin{eqnarray}\label{37}
{\cal{F}}_{\Psi}({k}_{1},{k}_{2},{k}_{3})=\frac{(2\pi)^4}{\prod_{i=1}^{3}{k}_i^3}{\cal{G}}_\Psi\,.
\end{eqnarray}
We note that, in solving the integral of equation (\ref{35}), we
have used the approximation that the coefficients in the brackets
of the lagrangian (\ref{32}) to be constants, because these
coefficients would vary slower than the scale factor. Furthermore,
the parameter ${\cal{G}}_{\Psi}$ is defined as
\begin{eqnarray}\label{38}
{\cal{G}}_{\Psi}=
\frac{3}{4}\bigg(1-\frac{1}{c_{s}^{2}}\bigg)\,{\cal{S}}_{1}
+\frac{1}{4}\bigg(1-\frac{1}{c_{s}^{2}}\bigg)\,{\cal{S}}_{2}\hspace{2cm}\nonumber\\
+\frac{3}{2M_{pl}}\bigg(\frac{1}{c_{s}^{2}}-1-\frac{2\lambda}{\Sigma}\bigg)
\,{\cal{S}}_{3}\,.\hspace{0.75cm}
\end{eqnarray}
in which we have the following relations for the shape functions
${\cal{S}}_{1}$, ${\cal{S}}_{2}$ and ${\cal{S}}_{3}$ respectively
\begin{equation}\label{39}
{\cal{S}}_{1}=\frac{2}{K}\sum_{i>j}k_{i}^{2}k_{j}^{2}-
\frac{1}{K^{2}}\sum_{i\neq j}k_{i}^{2}k_{j}^{3}\,,
\end{equation}

\begin{equation}\label{40}
{\cal{S}}_{2}=\frac{1}{2}\sum_{i}k_{i}^{3}
+\frac{2}{K}\sum_{i>j}k_{i}^{2}k_{j}^{2}-\frac{1}{K^{2}}
\sum_{i\neq j}k_{i}^{2}k_{j}^{3}\,,
\end{equation}

\begin{equation}\label{41}
{\cal{S}}_{3}= \frac{\big(k_{1}k_{2}k_{3}\big)^{2}}{K^{3}}\,,
\end{equation}
and also
\begin{eqnarray}\label{42}
K=\sum_{i} k_{i}\,.
\end{eqnarray}

It is obvious from equation (\ref{36}) that the three-point
correlator depends on the three momenta $k_1$, $k_2$ and $k_3$.
There are several different shapes of non-Gaussianities depending
on these wave numbers satisfying the condition $k_1 + k_2 + k_3 =
0$ \cite{65,66,67,68,69}. The simplest one is the so-called local
shape \cite{70,71,72,73}, which has a peak in the squeezed limit
$(k_3\rightarrow 0$ and $k_1\simeq k_2)$. The second shape
corresponds to the equilateral configuration \cite{74} with a
signal at $k_1 = k_2 = k_3$. There is another shape whose scalar
product with the equilateral template vanishes and is called the
orthogonal configuration \cite{75}. A linear combination of the
equilateral and orthogonal templates gives a shape corresponding
to folded triangle \cite{76} with a maximal signal in $k_1 = 2k_2
= 2k_3$ limit. We also note that, the orthogonal configuration has
a signal with a positive peak at the equilateral configuration and
a negative peak at the folded configuration. From the bispectrum
${\cal{G}}_{\Psi}$ of the three-point correlation function of
curvature perturbations, the non-linear parameter characterizing
the amplitude of non-Gaussianities is expressed by
\begin{equation}\label{43}
f_{_{NL}}=\frac{10}{3}\frac{{\cal{G}}_{\Psi}}{\sum_{i=1}^{3}k_{i}^{3}}\,.
\end{equation}
As has been stated previously, purely adiabatic Gaussian
perturbations give $f_{NL} = 0$, however, the presence of
non-Gaussian perturbations results in deviation from $f_{NL} = 0$.
Here we study the amplitude of non-Gaussianity in the equilateral
and orthogonal configurations. To this end, we should find the
bispectrum ${\cal{G}}_{\Psi}$ in these configurations. In this
regard, we follow \cite{77,78,79} and introduce a shape
${\cal{S}}_{*}^{equil}$ as
\begin{equation}\label{44}
{\cal{S}}_{*}^{equil}=-\frac{12}{13}\Big(3{\cal{S}}_{1}-{\cal{S}}_{2}\Big)\,.
\end{equation}
Moreover, we define another shape which is exactly orthogonal to
${\cal{S}}_{*}^{equil}$, as follows
\begin{equation}\label{45}
{\cal{S}}_{*}^{ortho}=\frac{12}{14-13\beta}\Big[\beta\big(3{\cal{S}}_{1}-{\cal{S}}_{2}\big)+3{\cal{S}}_{1}-{\cal{S}}_{2}\Big]\,,
\end{equation}
where $\beta\simeq 1.1967996$. Finally, making use of these
relations, the leading-order bispectrum (\ref{38}) can be written
in terms of the equilateral basis, ${\cal{S}}_{*}^{equil}$, and
the orthogonal basis, ${\cal{S}}_{*}^{ortho}$, as
\begin{equation}\label{46}
{\cal{G}}_{\Psi}={\cal{C}}_{1}\,{\cal{S}}_{*}^{equil} +
{\cal{C}}_{2} \,{\cal{S}}_{*}^{ortho}\,,
\end{equation}
where ${\cal{C}}_{1}$ and ${\cal{C}}_{2}$ are coefficients which
determine the magnitudes of the three-point correlation function
coming from equilateral and orthogonal contributions,
respectively, and are defined as
\begin{equation}\label{47}
{\cal{C}}_{1}=\frac{13}{12}\Bigg[\frac{1}{24}\bigg(1-\frac{1}{c_{s}^{2}}\bigg)\bigg(2+3\beta\bigg)
+\frac{\lambda}{12\Sigma}\bigg(2-3\beta\bigg) \Bigg]\,,
\end{equation}
and
\begin{equation}\label{48}
{\cal{C}}_{2}=\frac{14-13\beta}{12}\Bigg[\frac{1}{8}\bigg(1-\frac{1}{c_{s}^{2}}\bigg)
-\frac{\lambda}{4\Sigma}\Bigg]\,.
\end{equation}
$\lambda$ and $\Sigma$ are defined by equations (\ref{33}) and
(\ref{34}), respectively. Making use of equations
(\ref{44})-(\ref{48}), and also by definition of the non-linearity
parameter (\ref{43}), one can obtain the following expressions for
amplitude of the non-Gaussianity in the equilateral and orthogonal
configurations respectively
\begin{eqnarray}\label{49}
f_{_{NL}}^{equil}=\bigg(\frac{130}{36\sum_{i=1}^{3}k_{i}^{3}}\bigg)
\Bigg[\frac{1}{24}\bigg(1-\frac{1}{c_{s}^{2}}\bigg)\bigg(2+3\beta\bigg)\hspace{1cm}\nonumber\\
+\frac{\lambda}{12\Sigma}\bigg(2-3\beta\bigg)
\Bigg]{\cal{S}}_{*}^{equil}\,,\hspace{1cm}
\end{eqnarray}
and
\begin{equation}\label{50}
f_{_{NL}}^{ortho}=\bigg(\frac{140-130\beta}{36\,\sum_{i=1}^{3}k_{i}^{3}}\bigg)\Bigg[\frac{1}{8}\bigg(1-\frac{1}{c_{s}^{2}}\bigg)
-\frac{\lambda}{4\Sigma}\Bigg]{\cal{S}}_{*}^{ortho}\,.
\end{equation}
As has been mentioned previously, the shape function in the
equilateral configuration has a peak in $k_1 = k_2 = k_3$ limit
and also, the orthogonal shape has a signal with a positive peak
at the equilateral configuration. Thus, the nonlinearity parameter
in both configurations can be rewritten as
\begin{eqnarray}\label{51}
f_{_{NL}}^{equil}=\frac{325}{18}\Bigg[\frac{1}{24}\bigg(\frac{1}{c_{s}^{2}}-1\bigg)\bigg(2+3\beta\bigg)\hspace{2cm}\nonumber\\
+\frac{\lambda}{12\Sigma}\bigg(2-3\beta\bigg)\Bigg]\,,\hspace{1cm}
\end{eqnarray}
and
\begin{equation}\label{52}
f_{_{NL}}^{ortho}=\frac{10}{9}\Big(\frac{65}{4}\beta+\frac{7}{6}\Big)\Bigg[\frac{1}{8}\bigg(1-\frac{1}{c_{s}^{2}}\bigg)
-\frac{\lambda}{4\Sigma}\Bigg]\,.
\end{equation}

Up to this point, we have obtained the main equations of the two
field inflation model. In the following section, we examine the
model in confrontation with Planck 2015 TT, TE, EE+low P and
Planck2015 TTT, EEE, TTE and EET joint dataset to see the
consistency of this model. We also obtain some
constraints on the model's parameters space in this treatment.

\section{Observational Constraints}

In previous sections we have calculated the primordial
fluctuations in both linear and non-linear orders. An inflationary
model is successful and viable if its perturbation parameters are
consistent with observational data. So, in what follows we find
some observational constraints on the parameters space of the
model in hand. To this end, we should firstly define the form of
the DBI function $f(\chi)$. Usually, $f(\chi)$, which is related
to the geometry of the throat, is given in terms of the warp
factor of the AdS-like throat. In the pure AdS$_{5}$, this
function takes a simple form as $f(\chi)=\frac{\lambda}{\chi^4}$
\cite{47}.

We also emphasize that our method is based on the first-order
Hamilton-Jacobi formalism. As we have mentioned, slow-roll is not
the only possibility for successfully implementing models of
inflation, and solutions beyond the slow-roll approximation have
been found in particular situations. In fact, inflation is defined
to be a period of accelerated expansion, $(\frac{\ddot{a}}{a}>0)$,
indicating an equation of state in which vacuum energy dominates
over the kinetic energy of the field(s). In the slow-roll limit
the expansion of the universe is of the de Sitter form, with the
scale factor increasing exponentially in time $(H\simeq const,$
and $a \propto e^{Ht})$. With constant Hubble distance and
exponentially increasing scale factor, comoving length scales that
are initially smaller than the horizon, get rapidly redshifted
toward outside the horizon. In general, the Hubble parameter $H$
is not exactly a constant, but it varies slowly as the field(s)
evolve along the potential $V$. A convenient approach to the more
general case is to express the Hubble parameter directly as a
function of the field(s) instead of as a function of time
\cite{59,80,81,82,83}. Thus we continue our analysis by
reformulating the equations of motion as first order
Hamilton-Jacobi equations, following \cite{20}. We concentrate on
solutions satisfying the following sum separable Hubble parameter
\begin{equation}\label{53}
H=H_0+H_1\phi+H_2\chi\,.
\end{equation}
As has been suggested by Kinney \cite{59,Tzi08}, as
long as $t$ is a single-valued function of fields separately, one
can express the Hubble parameter directly as a function of the
fields instead of as a function of time. In this case one can use
the Hamilton-Jacobi formalism to describe dynamics of inflation.
For this reasons, we use the ansatz (\ref{53}) in our setup (see
\cite{Byr09,Eme12} for other occasions of the sum
separable Hubble parameter in literature).

Focussing on a homogeneous universe, the equations of motion for
both ordinary and DBI field can be written as a first order
Hamilton-Jacobi system as (see Appendix A) 
\begin{equation}\label{54}
\dot{\phi}=-2\frac{\partial H}{\partial \phi}
\end{equation}
\begin{equation}\label{55}
\dot{\chi}=-2c_{\chi}\frac{\partial H}{\partial \chi}
\end{equation}
where $c_{\chi}$ is the individual sound speed of the DBI field
and is given by
\begin{equation}\label{56}
c_{\chi}=\frac{1}{\sqrt{1+4f {H_{,\chi}^2}}}\,.
\end{equation}
The above Hamilton-Jacobi equations of motion have the solutions
as follows (see Appendix B)
\begin{equation}\label{57}
\phi(t)=-2H_1t+\phi_{0}
\end{equation}
\begin{equation}\label{58}
\chi(t)\simeq\frac{\chi_{0}\sqrt{\lambda}}{\chi_{0}t+\sqrt{\lambda}}
\end{equation}
leading to the scale factor as
\begin{eqnarray}\label{59}
a(t)=a_0\exp\Big((H_0+H_1\phi_0)t\hspace{3cm}\nonumber\\
-H_1^2t^2+H_2\sqrt{\lambda}\ln(\chi_0t+\sqrt{\lambda})\Big)\,,\hspace{0.75cm}
\end{eqnarray}
where $\phi_0$ and $\chi_0$ correspond to the value of each fields
at $t_{*}$ (the time at which observable scales exit the horizon).
Furthermore, one can easily obtain the corresponding potential of
the scalar fields by using the friedmann equation (\ref{5})
\begin{eqnarray}\label{60}
V(\phi,\chi)= 3(H_{0}+H_{1}\phi+H_{2}\chi)^{2}-2H_{1}^{2}\hspace{2cm}\nonumber\\
-\frac{\chi^{4}}{\lambda}\Bigg[\Bigg(1-\frac{4\lambda c_{\chi}^{2}
H_{2}^{2}}{\chi^{4}}\Bigg)^{-1/2}-1\Bigg]\,.\hspace{0.75cm}
\end{eqnarray}

After deriving the form of the potential, we explore the behavior
of the tensor-to-scalar ratio versus the scalar spectral index and
the orthogonal configuration versus the equilateral configuration
in the background of the Planck2015 TTT, EEE, TTE and EET data in
order to see the viability of this theoretical model in
confrontation with the recent observations. We note that in our
numerical study the values of $H_{0}$, $H_{1}$ and $H_{2}$ are
chosen so that for a defined number of e-folds, the inflation
phase terminates gracefully. For instance we have set
$H_{0}=10^{-6}$, $H_{1}=0.01$ and $H_{2}=0.01$. Note also that we
considered the case where inflation is initially driven dominantly
by the ordinary scalar field $\phi$. The initial values for the
two fields are $\phi_{0}=10$ and $\chi_{0}=1$.

\begin{figure*}
\flushleft\leftskip0em{
\includegraphics[width=.38\textwidth,origin=c,angle=0]{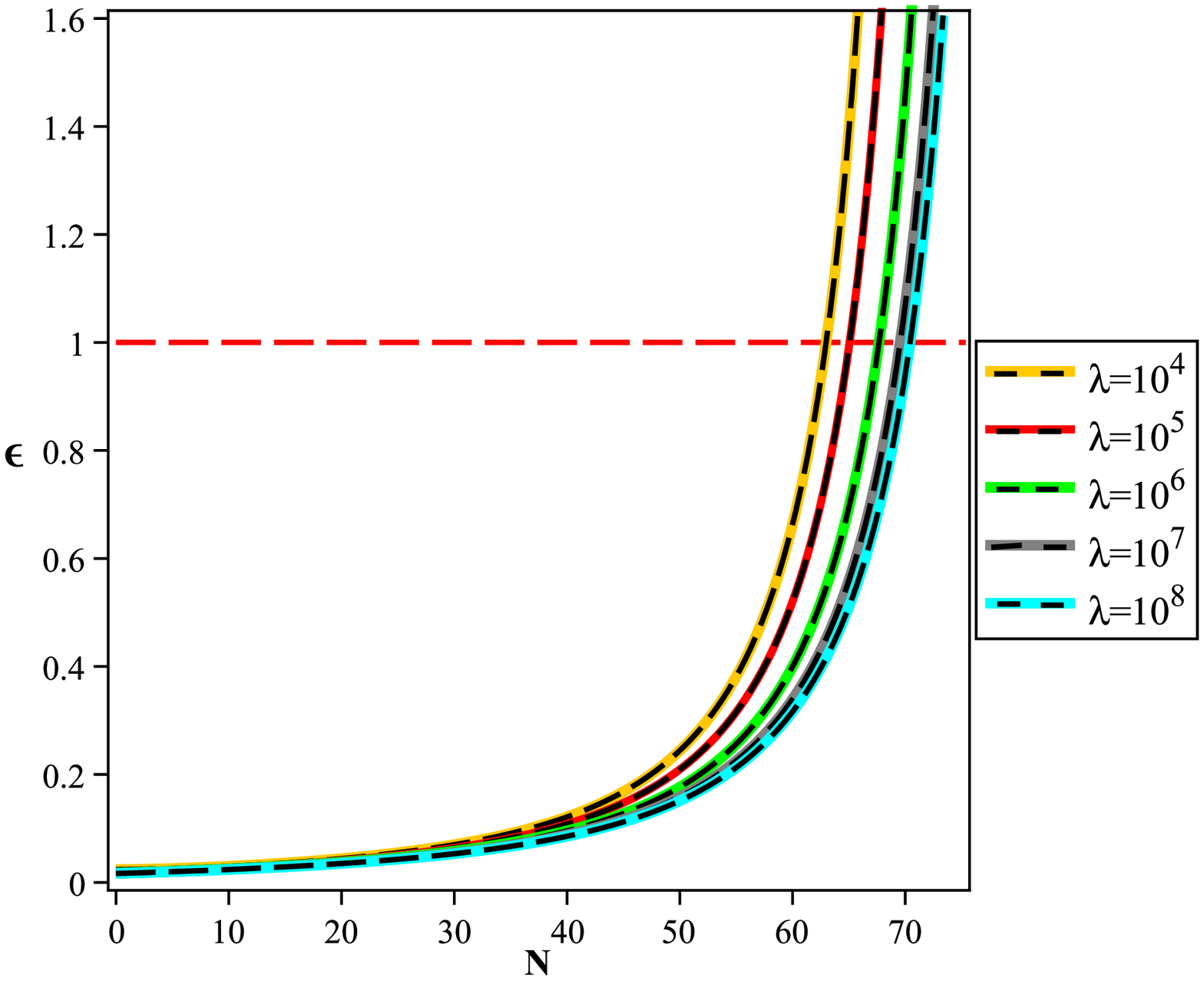}
\hspace{2cm}
\includegraphics[width=.42\textwidth,origin=c,angle=0]{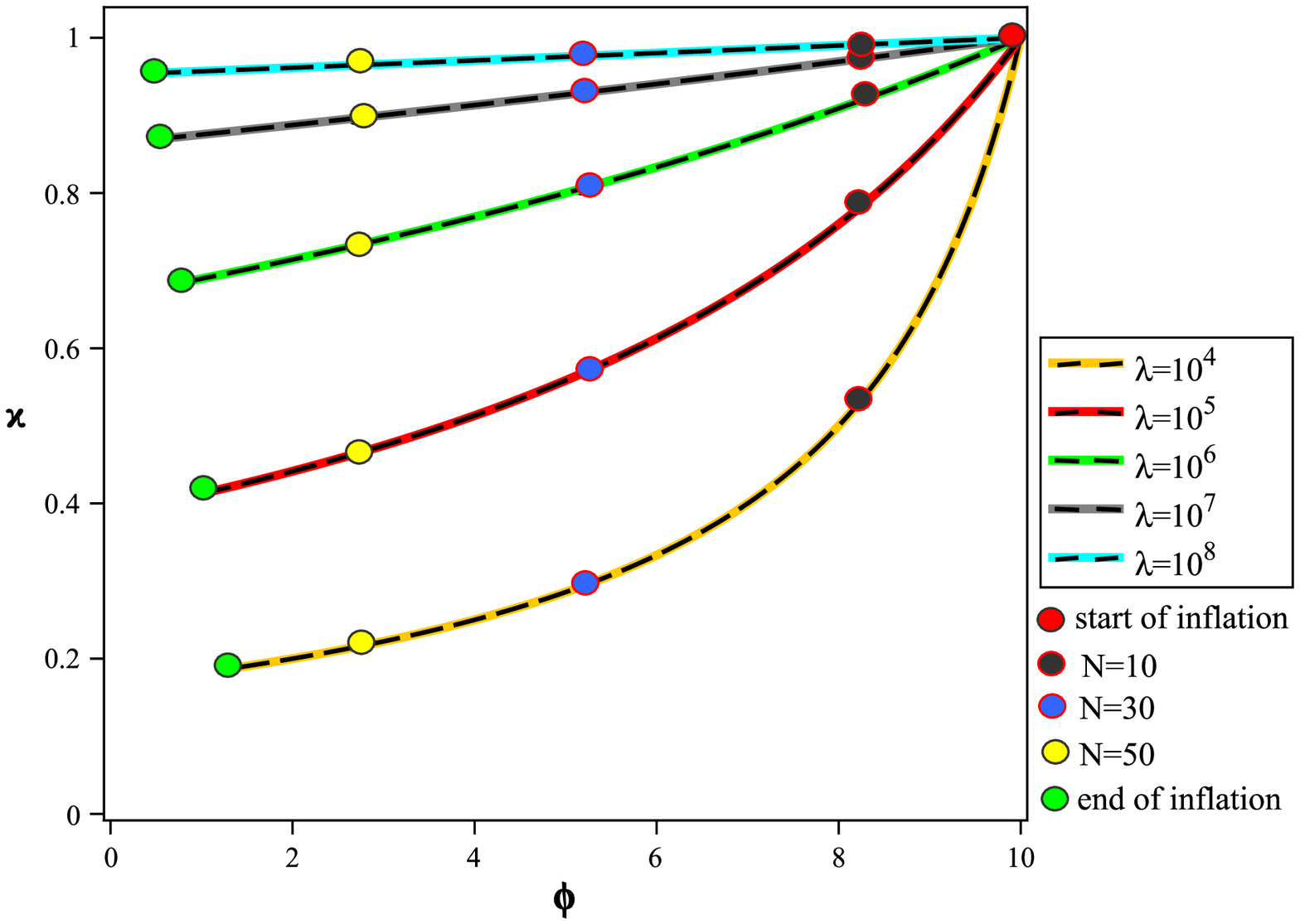}}
\caption{\label{fig:1} Left Panel: Variation of $\epsilon$ versus
$N$ for different values of $\lambda$. Right Panel: Trajectories
in the fields space, originating at $\phi_0=10$ and $\chi_0=1$ and
ending at some values on $N$ around $N \simeq 65$ e-folds of
inflation when $\epsilon = 1$.}
\end{figure*}

\begin{figure*}
\flushleft\leftskip0em{
\includegraphics[width=.38\textwidth,origin=c,angle=0]{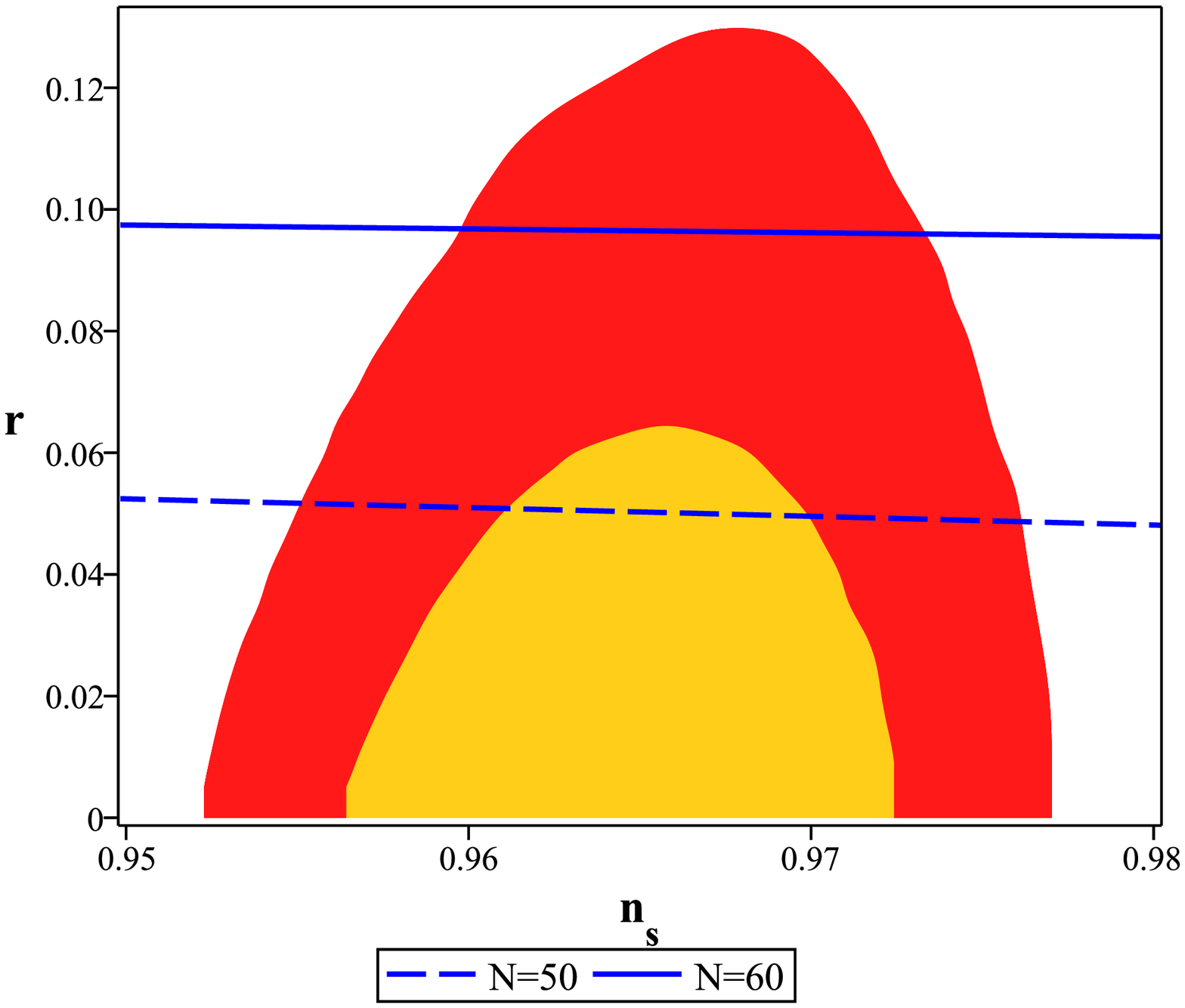}
\hspace{2cm}
\includegraphics[width=.40\textwidth,origin=c,angle=0]{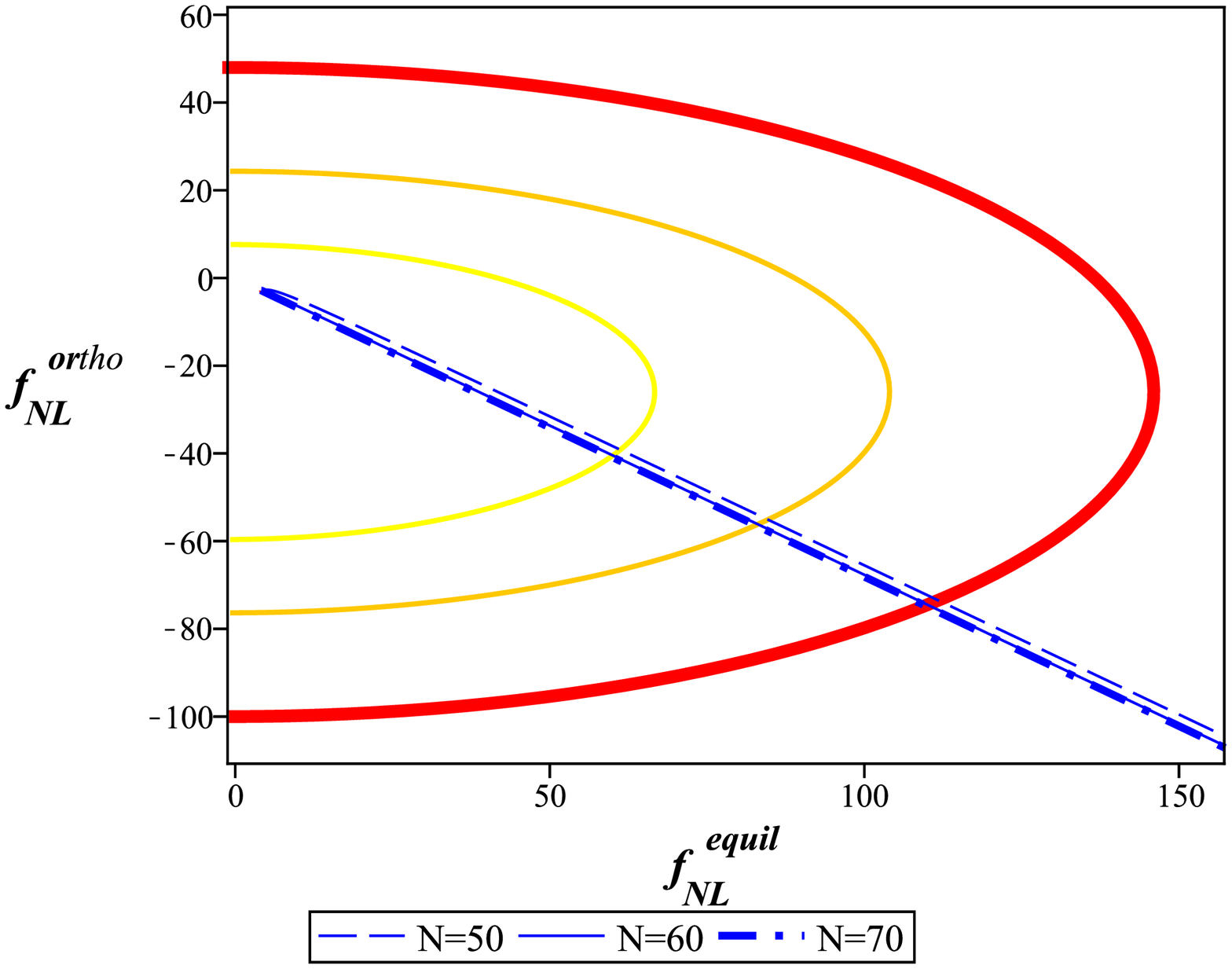}}
\caption{\label{fig:2} Tensor-to-scalar ratio versus the scalar
spectral index in the background of Planck2015 TT, TE, EE+low P
data (left panel), and the amplitude of the orthogonal versus
equilateral configuration of non-Gaussianity in the background of
Planck2015 TTT, EEE, TTE and EET data (right panel). Note that
these figures are plotted with $N=50, 60$ and $70$, for the
geometric function of the DBI field as ${\lambda}/{\chi^4}$.}
\end{figure*}

Fig.~\ref{fig:1} in the left panel shows variation of slow-roll
parameter $\epsilon$ versus $N$ for different values of $\lambda$.
For instance, the condition $\epsilon=1$ with $\lambda=10^{6}$ is
achieved for $N=67$. In the right panel of this figure, the
trajectories in the fields space are depicted. For relatively
small values of $\lambda$, the field $\phi$ plays the central
role. For large values of $\lambda$ the DBI field takes more
important role in the dynamics of inflation relative to its role
in the case with smaller $\lambda$. The left panel of
Fig.~\ref{fig:2} shows the behavior of the tensor-to-scalar ratio
versus the scalar spectral index for inflationary model with two
scalar fields (an ordinary field and a DBI field) for $N$ = 50, 60
and 70. Our numerical analysis shows that, although this model is
not consistent with Planck2015 dataset for $N=70$, but it will be
consistent with observation, if $20400<\lambda<47000$ for $N=50$
and $88200<\lambda<95000$ for $N=60$. The behavior of the
orthogonal configuration versus the equilateral non-Gaussianity of
this model is shown in the right panel of Fig.~\ref{fig:2}. This
figure confirms that in some ranges of the geometric parameter of
the DBI field, that is, $\lambda$, it is possible to have large
non-Gaussianity. For instance, for $N=60$ and $\lambda<911800$,
large non-Gaussianity can be realized in this setup. The ranges of
$\lambda$ in which the values of the inflationary parameters $r$
and $n_s$ and also, $f^{ortho}_{NL}$ and $f^{equi}_{NL}$, are
compatible with the $95\%$ confidence level of the Planck2015 TT,
TE, EE+low P and Planck2015 TTT, EEE, TTE and EET joint dataset
respectively are shown in Table~\ref{tab:1}.

\begin{table*}
\begin{center}
\caption{\label{tab:1}The ranges of $\lambda$ in which the values
of the inflationary parameters $r$ and $n_s$ and also,
$f^{ortho}_{NL}$ and $f^{equi}_{NL}$ are compatible with the
$95\%$ CL of the Planck2015 TT, TE, EE+ low P and Planck2015 TTT,
EEE, TTE and EET joint dataset respectively.}
\begin{tabular}{ccccc}
\\ \hline \hline$N$& \,\,\,\,\,\,& \,\,\,\,\,\,  & $r$\, versus \,$n_s$&$f^{ortho}_{NL}$\, versus\,$f^{equi}_{NL}$\\ \hline\\
$50$& \,\,\,\,\,\,& \,\,\,\,\,\,  &$20400<\lambda<47000$&$\lambda<2102800$\\\\
$60$& \,\,\,\,\,\, & \,\,\,\,\,\,  & $88200<\lambda<95000$ & $\lambda<911800$\\\\
$70$& \,\,\,\,\,\,& \,\,\,\,\,\,  &$\,not\,\, consistent\,$&$\lambda<357260$\\ \hline\\\\
\end{tabular}
\end{center}
\end{table*}

Now we derive the form of the scale factor versus the involving fields in order
to depict evolution in the fields space. Since $\frac{da}{a}=Hdt$, using the sum separable Hubble parameter
(\ref{53}), it can be written as
\begin{eqnarray}\label{77}
\frac{da}{a}=(H_{0}+H_{1}\phi+H_{2}\chi)dt\,.
\end{eqnarray}
After integration we have
\begin{eqnarray}\label{78}
\ln{\frac{a}{a_{0}}}=\int{H_{0}}dt+\int{\frac{H_{1}\phi}{\dot{\phi}}}d\phi+\int{\frac{H_{2}\chi}{\dot{\chi}}}d\chi\,,
\end{eqnarray}
which results in the following relation
\begin{eqnarray}\label{79}
a(\phi,\chi)=a_{0}\exp{\Big\{\frac{1}{4}(\phi_{0}^{2}-\phi^{2})-\frac{H_{0}}{2H_1}
(\phi-\phi_{0})\Big\}}\times\hspace{1.5cm}\\\nonumber
\exp{\Big\{\frac{1}{2}\sqrt{\lambda
H_2^2}\ln\Big(\frac{4}{\chi^{2}}\Big(2\lambda
H_{2}^{2}+\sqrt{\lambda H_{2}^{2}(\chi^{4}+4\lambda
H_{2}^{2})}\Big)\Big)\Big\}}\times\\\nonumber
\exp{\Big\{\frac{1}{4}\Big(\sqrt{\chi_{0}^{4}+4\lambda
H_{2}^{2}}-\sqrt{\chi^{4}+4\lambda
H_{2}^{2}}\Big)\Big\}}\times\hspace{1.2cm}\\\nonumber
\exp{\Big\{-\frac{1}{2}\sqrt{\lambda
H_2^2}\ln\Big(\frac{4}{\chi_{0}^{2}}\Big(2\lambda
H_{2}^{2}+\sqrt{\lambda H_{2}^{2}(\chi_{0}^{4}+4\lambda
H_{2}^{2})}\Big)\Big)\Big\}}\,.
\end{eqnarray}
Figure 3 shows the evolution in fields space for $\lambda=10^{5}$.\\

\begin{figure}[htp]
\begin{center}
\includegraphics{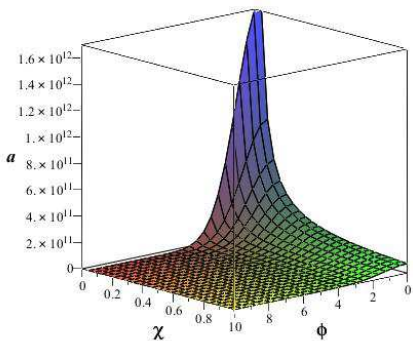}\vspace{5cm}
\end{center}
\caption{\small {Evolution in fields space for $\lambda=10^{5}$.}}
\end{figure}

\section{Conclusion}

In this paper we have studied the dynamics of an inflationary
model driven by two scalar fields, an ordinary scalar field with
canonical kinetic term and a DBI field with non-canonical kinetic
term. At first, we have obtained the main equations of the model.
Then, we have studied the linear perturbations of this
inflationary model using the ADM formalism. By expanding the
action of the model up to the second order in perturbation, we
have derived the two-point correlation functions which result in
the amplitude of the scalar perturbation and its spectral index.
Also, by studying the tensor part of the perturbed metric, we have
obtained the tensor perturbation and its spectral index as well.
The ratio between the amplitude of the tensor and scalar
perturbations has been obtained in this setup. In order to study
the non-Gaussian feature of the primordial perturbations in this
setup, we have studied the non-linear theory in details. To
investigate non-linear perturbation in the model, one has to
expand the action up to the cubic order in perturbation and
calculate the three-point correlation functions. Thus, by using
the interacting picture we have computed the three-point
correlation functions and the nonlinearity parameter in our setup.
By introducing the shape functions as ${\cal{S}}_{*}^{equil}$ and
${\cal{S}}_{*}^{ortho}$, we have obtained the amplitude of the
non-Gaussianity in the equilateral and orthogonal configurations.
We have focused in the limit $k_{1} = k_{2} = k_{3}$, in which,
both the equilateral and orthogonal configuration have peak.

After calculating the main perturbation parameters, we have tested
our model with recent observational data. Note that we have worked
beyond the slow-roll approximation, but we have assumed a
separable Hubble parameter. In other words, since slow-roll
condition can be temporarily violated during inflation, we have
gone beyond this approximation and our method is based on the
first order Hamilton-Jacobi formalism, which allows us to express
inflationary parameters in the model, without having to focus on a
slow-roll regime. We have also defined the form of the DBI
function, $f(\chi)$, in terms of the warp factor of the AdS-like
throat as $f(\chi)=\frac{\lambda}{\chi^4}$. Then, we have studied
this inflationary model numerically and compared our model with
the recently released observational data. To this end, we have
studied the behavior of the tensor-to-scalar ratio versus the
scalar spectral index in the background of the Planck2015 TT, TE,
EE+low P data and obtained some constraints on model's parameters
space. Furthermore, by studying the behavior of the orthogonal
configuration versus the equilateral configuration in the
equilateral limit and in the background of the Planck2015 TTT,
EEE, TTE and EET data, the non-Gaussinaty feature of the
primordial perturbations have been analyzed numerically. In this
paper, we have shown that this inflationary model is
observationally viable in some ranges of the DBI parameter. As an
important result, we have shown that this models allows to have
large non-Gaussianity that would be observable by future
improvements in experiments. On the other hand, the trajectories
in the fields space have been depicted which indicate that for
relatively small values of $\lambda$, the field $\phi$ plays the
central role in deriving inflation. However, for large values of
$\lambda$, the DBI field takes more important role in the dynamics
of inflation relative to its role in the case with smaller
$\lambda$.\\

{\bf Appendix A: Equations of motion in
Hamilton-Jacobi formalism}\\

In this appendix we obtain the equations of motion (\ref{3}) and
(\ref{4}) in Hamilton-Jacobi formalism. For an ordinary scalar
field the Friedmann equation and the equation of motion
are
\begin{eqnarray}
H^{2}=\frac{1}{3}(\frac{1}{2}{\dot{\phi}}^{2}+V(\phi))\,,\label{61}
\end{eqnarray}
and 
\begin{eqnarray}
{\ddot{\phi}}+3H{\dot{\phi}}+V_{,\phi}=0\,,\label{62}
\end{eqnarray}
respectively. Differentiating equation (\ref{61}) with respect to time results in 
\begin{eqnarray}
2HH_{,\phi}{\dot{\phi}}=\frac{1}{3}{(\ddot{\phi}+V_{,\phi})}\dot{\phi}\,.\label{63}
\end{eqnarray}
Using the equation of motion (\ref{62}) we can easily simplify the
right-hand side of this relation as
\begin{eqnarray}
2HH_{,\phi}{\dot{\phi}}=-H\dot{\phi}^2\,,\label{64}
\end{eqnarray}
and by substituting back into the definition of the Hubble
parameter in the Friedmann equation, we find the following two
first-order equations which are entirely equivalent to the
second-order equation of motion (\ref{62})
\begin{eqnarray}
\dot{\phi}=-2H_{,\phi}\,,\label{65}
\end{eqnarray}
and
\begin{eqnarray}\label{66}
3H^2=2H_{,\phi}^2+V\,,
\end{eqnarray}
which are the Hamilton-Jacobi
equations. In our case with an ordinary
scalar field and a DBI field, by differentiating the Friedmann 
equation (\ref{5}) with respect to time, we obtain
\begin{eqnarray}\label{67}
2H{\dot{H}}=\frac{1}{3}\Bigg((\ddot{\phi}+V_{,\phi})\dot{\phi}+\hspace{4.5cm}\\\nonumber
\Big(\gamma^{3}\ddot{\chi}+\frac{1}{2}\gamma^{3}f_{,\chi}f^{-2}+f_{,\chi}f^{-2}-\frac{3}{2}\gamma
f_{,\chi}f^{-2}+V_{,\chi}\Big)\dot{\chi}\Bigg)\,.
\end{eqnarray}
We can also write
\begin{eqnarray}\label{68}
\dot{H}=H_{,\phi}\dot{\phi}+H_{,\chi}\dot{\chi}\,.
\end{eqnarray}
Finally using equations of motion (\ref{3}) and (\ref{4}) one can
easily find the following first order equations
\begin{eqnarray}\label{69}
\dot{\phi}=-2H_{,\phi}\,,
\end{eqnarray}
\begin{eqnarray}\label{70}
\dot{\chi}=-\frac{2}{\gamma}H_{,\chi}\,,
\end{eqnarray}
and also
\begin{eqnarray}\label{71}
3H^2=2H_{,\phi}^{2}+f^{-1}\Big(\frac{1}{c_{\chi}}-1\Big)+V\,.
\end{eqnarray}
We note that according to definition of $c_{\chi}$ in equation
(\ref{56}) it is obvious that $c_{\chi}=\gamma^{-1}$ and we have therefore 
\begin{eqnarray}\label{72}
\dot{\chi}=-2c_{\chi}H_{,\chi}\,.
\end{eqnarray}

{\bf Appendix B: Derivation of Eq. (\ref{58})}\\

In order to derive equation (\ref{58}), we note that since
\begin{equation}\label{73}
\dot{\chi}=-2c_{\chi}\frac{\partial H}{\partial \chi}\,,
\end{equation}
one should solve the following integral to find $\chi(t)$ 
\begin{equation}\label{74}
\int{dt}=\frac{-1}{2H_{2}}\int{\sqrt{1+\frac{4\lambda
H_{2}^{2}}{\chi^{4}}}}d{\chi}\,.
\end{equation}
The right hand side integral cannot be solved analytically. To find an approximate analytical solution, 
we can neglect the unity in comparison with the second term in the square root (this is reasonable at least for sufficiently large $\lambda$)  
 
\begin{equation}\label{75}
\int{dt}\simeq\frac{-1}{2H_{2}}\int{\sqrt{\frac{4\lambda
H_{2}^{2}}{\chi^{4}}}}d{\chi}\,,
\end{equation}
and it follows
\begin{equation}\label{76}
\frac{1}{\chi(t)}-\frac{1}{\chi_{0}}\simeq\frac{t}{\sqrt{\lambda}}\,.
\end{equation}

{\bf Acknowledgement}
We would like to thank Dr Narges Rashidi for fruitful discussions. 
We appreciate also an anonymous referee for insightful comments which
considerably improved the quality of the paper.

\end{document}